\documentclass{svproc}
\usepackage{url}
\usepackage{mathtools}
\usepackage{overpic}
\newcommand{\bb}{\mathbf b}
\newcommand{\bu}{\mathbf u}
\newcommand{\bk}{\mathbf k}

\newcommand{\bm}[1]{\mbox{\boldmath{$#1$}}}
\newtheorem{rem}{Remark}
\def\cl {\nonumber \\}

\def\0{\boldsymbol{0}}

\def\grad{\nabla}

\begin{document}
\include{commands.tex}

\mainmatter              
\title{GEA: a new finite volume-based open source code for the numerical simulation of atmospheric and ocean flows}
\titlerunning{GEA}  
%
\author{Michele Girfoglio\inst{1}, Annalisa Quaini\inst{2} \and Gianluigi Rozza\inst{1}}
\authorrunning{Michele Girfoglio et al.} 
%
\tocauthor{Michele Girfoglio, Annalisa Quaini and Gianluigi Rozza}
\institute{mathLab, Mathematics Area, SISSA, via Bonomea 265, I-34136 Trieste, Italy, \\
\email{michele.girfoglio@sissa.it}, \email{gianluigi.rozza@sissa.it}
\and
Department of Mathematics, University of Houston, Houston TX 77204, USA, \email{quaini@math.uh.edu}}

\maketitle              

\begin{abstract}
We introduce GEA (Geophysical and Environmental Applications), a new open-source atmosphere and ocean modeling framework within the finite volume C++ library OpenFOAM\textsuperscript{\textregistered}. 
Here, we present the development of a non-hydrostatic atmospheric model consisting of a pressure-based solver for the Euler equations written in conservative form using density, momentum, and total energy as variables. We validate the solver for two idealized test cases involving buoyancy driven flows: smooth and non-smooth rising thermal bubble. 
Through qualitative and quantitative comparisons against numerical data available in the literature, we show that our approach is accurate. 
\end{abstract}
\section{Introduction}
Fast and accurate weather/climate forecasts need state-of-the-art numerical and computational methodologies. Open
source software packages for weather and climate simulations, i.e. the Climate Machine \cite{clima} and WRF \cite{WRF}, are very useful tools for realistic simulations, but testing and assessing new numerical approaches within them is non-trivial. 
This paper is a follow-up of our work presented in \cite{Girfoglio2023} and it is meant to further lay the foundation for a new open source package, called GEA (Geophysical and Environmental Applications) \cite{GEA}. GEA is specifically created for the assessment of new computational approaches for the simulation of mesoscale atmospheric flows and ocean flows \cite{GIRFOGLIO2023114656,GQR_ROM_QGE22,Girfoglio2023}. 
To maximize the reach and the impact, as well as to facilitate sharing it with the scientific community, we choose to build our software package
on OpenFOAM\textsuperscript{\textregistered} \cite{Weller1998}, an open source, freely available C++ finite volume library that has become
widely used in Computational Fluid Dynamics (CFD).

As the core of our open source package, we present a solver for the Euler
equations for non-hydrostatic mesoscale atmospheric
modeling, and assess it through two well-known test cases involving a smooth and non-smooth rising thermal bubble. We consider the Euler equations written
in conservative form using density, momentum, and total energy as variables.

The rest of the paper is organized as follows. In Sec. 2, we briefly describe the formulation of
the Euler equations under consideration. Sec. 3 presents our pressure-based
approach and provides the details of space and time discretization. Numerical results for the two
benchmark tests are discussed in Sec. 4. Conclusions are drawn in Sec. 5.

\section{Problem definition}
\subsection{The compressible Euler equations}
\label{sec:NS Equations}

We consider the dynamics of the dry atmosphere (i.e., no moisture) in a spatial domain of interest $\Omega$ by neglecting the effects of solar radiation and heat flux from the ground.
We assume that dry air behaves like an ideal gas.

Let $\rho$ be the air density, $\bu$ = $(u, v, w)$  
the wind velocity, and $e$ the total energy density. Note that 
$e = c_v T + |\bu|^2/2 + g z$, where $c_{v}$ the specific heat capacity at constant volume, $T$ is the absolute temperature, $g$ is the gravitational constant, and $z$ is the vertical coordinate.
The equations stating conservation of mass, momentum, and energy for the dry atmosphere written in terms of $\rho$, $\bu$, and $e$ over a time interval of interest $(0,t_f]$ read: 
\begin{align}
&\frac{\partial \rho}{\partial t} + \nabla \cdot (\rho \bu) = 0 &&\text{in } \Omega \times (0,t_f], \label{eq:mass}  \\
&\frac{\partial (\rho \bu)}{\partial t} +  \nabla \cdot (\rho \bu \otimes \bu) + \nabla p   + \rho g \widehat{\bk} = \boldsymbol{0} &&\text{in } \Omega \times (0,t_f],  \label{eq:mom} \\
&\frac{\partial (\rho e)}{\partial t} +  \nabla \cdot (\rho \bu e) + \nabla \cdot (p \bu) = 0 &&\text{in } \Omega \times (0,t_f],
\label{eq:ent}
\end{align}
where $\widehat{\bk}$ is the unit vector aligned with the vertical axis $z$ and $p$ is pressure. To close system \eqref{eq:mass}-\eqref{eq:ent}, we need a 
thermodynamics equation of state for $p$. Following the assumption that dry air behaves like an ideal gas, we have: 
\begin{align}
p = \rho R T, 
\label{eq:p}
\end{align}
where $R$ is the specific gas constant of dry air.

Let us write the pressure as the sum of a fluctuation $p'$ with respect to a background state
\begin{align}
p = p' + \rho g z. \label{eq:p_split}
\end{align}
By plugging \eqref{eq:p_split} into \eqref{eq:mom}, we obtain:
\begin{align}
\frac{\partial (\rho \bu)}{\partial t} +  \nabla \cdot (\rho \bu \otimes \bu) + \nabla p' + gz \nabla \rho = \boldsymbol{0} \quad \text{in } \Omega \times (0,t_f].  \label{eq:mom_split}
\end{align}




Let $c_{p}$ be the specific heat capacity at constant pressure for dry air and let
\begin{equation}\label{eq:K_h}
K = |\bu|^2/2, \quad h = c_v T + p/\rho = c_p T,    
\end{equation}
be the kinetic energy density and the specific enthalpy, respectively. The total energy density can be written as $e = h - p/\rho + K + gz$. Then, eq.~\eqref{eq:ent} can be rewritten as:
\begin{align}
\frac{\partial (\rho h)}{\partial t} +  \nabla \cdot (\rho \bu h) + 
\frac{\partial (\rho K)}{\partial t} +  \nabla \cdot (\rho \bu K) - \dfrac{\partial p}{\partial t}  +  
\rho g \bu \cdot \widehat{\bk} = 0,
\label{eq:over_ent}
\end{align}
where we have used eq.~\eqref{eq:mass} for further simplification.  

\begin{rem}
To preserve the numerical stability, we add an artificial diffusion term to eq.~\eqref{eq:mom_split} and \eqref{eq:over_ent}:
\begin{align}
&\frac{\partial (\rho \bu)}{\partial t} +  \nabla \cdot (\rho \bu \otimes \bu) + \nabla p' + gz \nabla \rho -  \mu \Delta \bu = \boldsymbol{0} \label{eq:mom_LES}  \\
&\frac{\partial (\rho h)}{\partial t} +  \nabla \cdot (\rho \bu h) + 
\frac{\partial (\rho K)}{\partial t} +  \nabla \cdot (\rho \bu K) - \dfrac{\partial p}{\partial t}  +  
\rho g \bu \cdot \widehat{\bk}  -  \frac{\mu_a}{Pr} \Delta h = 0,
\label{eq:ent_LES}
\end{align}
where $\mu$ is a constant diffusivity coefficient and $Pr$ is the Prandtl number.
\end{rem}\label{rem:1}

\begin{rem}
A quantity of interest for atmospheric problems is the potential temperature $\theta$ defined as
\begin{align}
\theta = \frac{T}{\pi}, \quad \pi = \left( \frac{p}{p_0} \right)^{\frac{R}{c_{p}}}, \label{eq:theta}
\end{align}
where $p_0 = 10^5$ Pa is the atmospheric pressure at the ground. Additionally, 
we define the potential temperature fluctuation $\theta'$:
\begin{align}
\theta'(x,y,z,t) = 
\theta(x,y,z,t) - \theta_0(z), \label{eq:theta_split} 
\end{align}
where $\theta_0$ is the mean hydrostatic value, which is a  function
of the vertical coordinate $z$ only. 
\end{rem}

 We will devise a splitting approach for problem \eqref{eq:p}-\eqref{eq:p_split},\eqref{eq:K_h},\eqref{eq:mom_LES}-\eqref{eq:ent_LES}, because this formulation of the Euler equation facilitates the decoupling of all variables and it allows for an explicit treatment of the kinetic and potential energies.

\section{Time and space discretization}
This section briefly presents a space and time discretization for the model
\eqref{eq:mass},\eqref{eq:p}-\eqref{eq:p_split},\eqref{eq:K_h},\eqref{eq:mom_LES}-\eqref{eq:ent_LES}.
For the space discretization,
we adopt a finite volume method. This requires to partition the computational domain $\Omega$ into cells or control volumes $\Omega_i$, with $i = 1, \dots, N_{c}$, where $N_{c}$ is the total number of cells in the mesh. Let  \textbf{A}$_j$ be the surface vector of each face of the control volume, 
with $j = 1, \dots, M$. With the subindex $i$ we will denote a variable average in control volume $\Omega_i$.
Let $\Delta t \in R$, 
$t^n = n \Delta t$, with $n = 0, ..., N_f$ and $t_f = N_f \Delta t$. Moreover, we denote by $y^n$ the approximation of a generic quantity $y$ at the time $t^n$. For time discretization, we adopt a Backward Differentiation Formula of order 1 (BDF1). A monolithic approach for coupled problem \eqref{eq:p}-\eqref{eq:p_split},\eqref{eq:K_h},\eqref{eq:mom_LES}-\eqref{eq:ent_LES} would lead to high computational costs. 
Thus, to save computational time we adopt a splitting approach consisting of three steps detailed below.

Problem \eqref{eq:p}-\eqref{eq:p_split},\eqref{eq:K_h},\eqref{eq:mom_LES}-\eqref{eq:ent_LES}
discretized in time and space reads: given $\rho^0$, $\bu^0$, $h^0$,  $p^0$, and $T^0$, set $K^0 = |\bu^0|^2/2$ and for $n \geq 0$ find solution $(\rho^{n+1}_i, \bu^{n+1}_i,h^{n+1}_i,K^{n+1}_i, \\p^{n+1}_i, p'^{,n+1}_i, T^{n+1}_i)$ of system: 

\begin{itemize}
\item[-] \emph{Step 1}: find first intermediate density ${\rho}_i^{n+\frac{1}{3}}$, intermediate velocity ${\bu}_i^{n+\frac{1}{3}}$ and associated kinetic energy density $K_i^{n+\frac{1}{3}}$ such that
\begin{align}
&\frac{1}{\Delta t} {\rho}^{n+\frac{1}{3}}_i + \sum_j  \varphi^n_j = b_{\rho,i}^{n+1}, \quad \varphi^n_j =(\rho^n\bu^{n})_{i,j}  \cdot  \textbf{A}_j,\label{eq:e1_d} \\
& \frac{1}{\Delta t} {\rho}^{n+\frac{1}{3}}_i \bu^{n+\frac{1}{3}}_i +\sum_j^{} \varphi^n_j \bu^{n+\frac{1}{3}}_{i,j} + \nabla p^{n}_{i} - \sum_j^{} \mu \nabla (\bu^{n+\frac{1}{3}}_i)_j \cdot \textbf{A}_j = {\bm b}^{n+1}_{\bu, i}, \label{eq:step2_sd} \\
& K^{n+\frac{1}{3}}_i = \frac{|\bu_i^{n+\frac{1}{3}}|^2}{2}, \label{eq:step1_3sd}
\end{align}
where $ \varphi^n_j$ denotes 
the convective
flux through face $j$ of $\Omega_i$, $b^{n+1}_{\rho,i} = \rho^n_i/\Delta t$ and $\bb^{n+1}_{\bu,i} = \rho^{n}_i\bu^n_i/\Delta t$.
\item[-] \emph{Step 2}: find average specific enthalpy $h^{n+1}_i$, temperature $T^{n+1}_i$, and second intermediate density ${\rho}^{n+\frac{2}{3}}_i$ in control volume $\Omega_i$ such that
\begin{align}
&\frac{1}{\Delta t} {\rho}^{n+\frac{1}{3}}_i h^{n+1}_i + \sum_j^{} \varphi^n_j h^{n+1}_{i,j} - \sum_j^{} \frac{\mu}{Pr} (\nabla h^{n+1}_i)_j \cdot \textbf{A}_j  = \tilde{b}_{e,i}^{n} \cl
&\quad - 
\frac{1}{\Delta t} {\rho}^{n+\frac{1}{3}}_i K^{n+\frac{1}{3}}_i - \sum_j^{} \varphi^n_j K^{n+\frac{1}{3}}_{i,j} + \frac{1}{\Delta t} p_i^n - {\rho}^{n+\frac{1}{3}}_i g \bu^{n+\frac{1}{3}}_i \cdot \widehat{\bk}, \label{eq:step3_sd}\\
& h^{n+1}_i - c_p T^{n+1}_i =  h^{n}_i - c_p T^{n}_i,
\label{eq:step3_2sd} \\
&{\rho}^{n+\frac{2}{3}}_i R T^{n+1}_i = p^{n}_i, \label{eq:step3_3sd}
\end{align}
where $\tilde{b}_e^{n} = (\rho^n h^n + \rho^n K^{n-1} - p^{n-1}) / \Delta t$. 

\item[-] \emph{Step 3}: find end-of-step velocity $\bu^{n+1}$ and associated kinetic energy density $K^{n+1}$, pressure $p^{n+1}$ and pressure fluctuation $p'^{,n+1}$, and end-of-step density $\rho^{n+1}$ 
such that 
\begin{align}
&\frac{1}{\Delta t} {\rho}^{n+\frac{1}{3}}_i \bu^{n+1}_i +\sum_j^{} \varphi^n_j \bu^{n+1}_{i,j} + \nabla p'^{,n+1}_{i}  + g z_i \grad{{\rho}_i^{n+\frac{2}{3}}} \cl
&\quad - 
\sum_j^{} \mu (\nabla \bu^{n+1}_i)_j \cdot \textbf{A}_j = {\bm b}^{n+1}_{\bu, i}, \label{eq:step2_sd}
\end{align}
\begin{align}
& \sum_j {\rho}_j^{n+\frac{2}{3}}(\nabla p'^{,n+1}_i)_j \cdot \textbf{A}_j =  \sum_j  \frac{ {\rho}_j^{n+\frac{2}{3}} \Delta t}{{\rho}_j^{n+\frac{1}{3}}} \left(\mathbf{H}(\bu_i^{n+1})_j  - gz_j (\nabla {\rho}_i^{n+\frac{2}{3}})_j \right) \cdot \textbf{A}_j \cl 
& \quad - 
b_{\rho, i}^{n+1} + \dfrac{1}{\Delta t} {\rho}_i^{n+\frac{2}{3}} 
, \label{eq:p_prime_sd} \\
& p^{n+1}_i = p'^{,n+1}_i + {\rho}^{n+\frac{2}{3}}_i g z_i, \quad
K^{n+1}_i = \frac{|\bu_i^{n+1}|^2}{2}, \quad 
\rho^{n+1}_i = 
\frac{p_i^{n+1}}{R T^{n+1}_i},\label{eq:step4_4sd}
\end{align}
where $\mathbf{H}(\bu_i^{n+1}) = - \sum_j^{} \varphi^n_j \bu^{n+1}_{i,j}
+ \sum_j^{} \mu (\nabla \bu^{n+1}_i)_j \cdot \textbf{A}_j + {\bm b}^{n+1}_{\bu, i}$.
\end{itemize}

In order to decouple the computation of the pressure from the computation of the velocity, we use the PISO algorithm \cite{Weller1998}. We choose a second-order accurate scheme for the Laplacian and gradient terms and a fourth-order accurate scheme for the divergence term.
For
more details, we refer the reader to \cite{Girfoglio2023}.
\section{Numerical results}
We validate our solver with respect to two classic benchmarks, the smooth and non-smooth rising thermal bubble. Both test cases involve a perturbation of a neutrally stratified atmosphere with
uniform background potential temperature over a flat terrain. In both tests,
the computational domain in the $xz$-plane is $\Omega=[0, 1000]^2$ m$^2$ and the time interval of interest is $(0, 600]$ s. Impenetrable, free-slip boundary conditions are imposed on all walls.

Since these benchmarks do not have an exact solution, one can only have a
comparison with other numerical data available in the literature. 



\subsection{Smooth rising thermal bubble}

The initial density is given by \begin{align}
\rho^0 = \frac{p_g}{R \theta_0} \left(\frac{p}{p_g}\right)^{c_{v}/c_p}, \quad p = p_g \left( 1 - \frac{g z}{c_p \theta^0} \right)^{c_p/R}, \label{eq:rho_wb}
\end{align}
with 
$c_p = R + c_v$, $c_v = 715.5$ J/(Kg K), $R = 287$ J/(Kg K).
In \eqref{eq:rho_wb}, $\theta^0$ is the initial potential temperature, which is defined as:
\begin{equation}
\theta^0 = 300  + \frac{0.5}{2}\left[  1 + \cos\left(\frac{\pi r}{r_c}\right)\right] ~ \textrm{if $r\leq r_c$},\quad\theta^0 = 300
~ \textrm{otherwise},
\label{dcEqn1}
\end{equation}
where $r = \sqrt[]{\left(x-x_{c}\right)^{2} + \left(z-z_{c}\right)^{2}}$, with $r_c = 250~\mathrm{m}$ and $(x_c,z_c) = (500, 350)~\mathrm{m}$.
The initial velocity field is zero everywhere and the initial specific enthalpy is given by \begin{align}
h^{0} = c_p \theta^0 \left( \frac{p}{p_g} \right)^{\frac{R}{c_{p}}}.
\label{eq:e0}
\end{align}
Following \cite{Restelli1}, we use a mesh with uniform resolution $h = \Delta x = \Delta z = 5$ m. The time step is set to $\Delta t = 0.1$ s. 
Furthermore, we set $\mu = 0.15$ and $Pr = 1$.

\begin{figure}[htb]
\centering
 \begin{overpic}[width=0.525\textwidth]{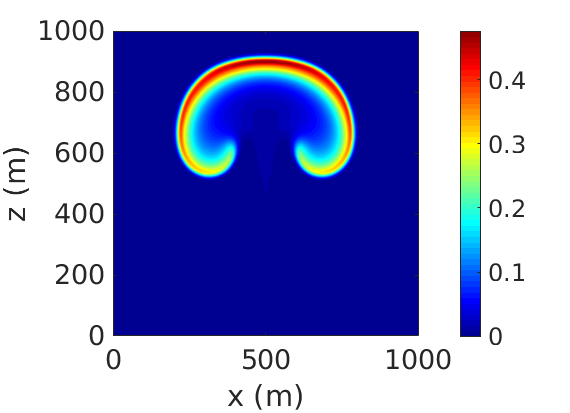}  
     \end{overpic}~ \hspace{0.05cm}
       \begin{overpic}[width=0.51\textwidth]{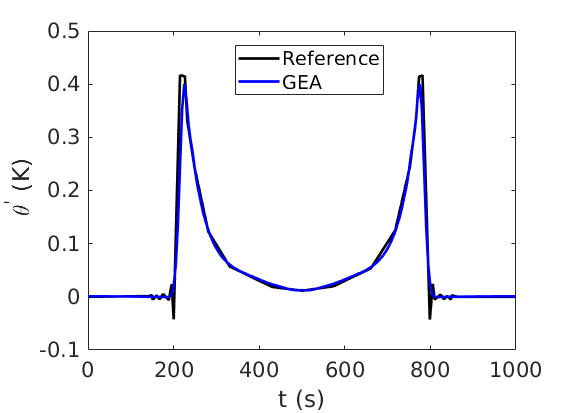}  
      \end{overpic}
\caption{Smooth rising thermal bubble: perturbation of potential
temperature computed at $t = 600$ s (left) and its profile at $z = 700$ m compared with data from \cite{Restelli1} denoted as ``Reference'' (right).}
\label{fig:RTB3}
\end{figure}

Figure \ref{fig:RTB3} (left) shows the potential temperature perturbation computed at $t = 600$ s. This plot is in very good agreement with the corresponding figure in \cite{Restelli1}. Figure \ref{fig:RTB3} (right) depicts the profile of the potential temperature perturbation along $z = 700$ m together with the data from \cite{Restelli1}. We observe that the two curves are very close, except for the oscillations at $t \approx 200$ s and $t \approx 800s$ that in our solution are completely damped and the slightly lower maxima. 

For further comparison, in Table \ref{tab:1} we report the extrema for
the horizontal velocity $u$ and vertical velocity $w$, together with the values obtained in \cite{Restelli1}.
The results from \cite{Restelli1} are obtained by using density-based approach developed from a Godunov-type scheme. Moreover, the authors of \cite{Restelli1} use a Discontinuous Galerkin method for the space discretization. 
Other differences with our methodology include the orders of space and time discretizations and the different treatment of the hydrostatic term. Given all these differences, we believe that our results are in line with the reference ones.


\begin{table}[htb]
\begin{center}
\begin{tabular}{|c|c|c|c|c|c|} \hline
Model & Res. [m] & $u_{min}$ (m/s) & $u_{max}$ (m/s) & $w_{min}$ (m/s) & $w_{max}$ (m/s) \\
 \hline
GEA & 5 & -1.898 & 1.898 & 1.682 & 2.495\\
Ref. \cite{Restelli1} & 5 & -2.161 & 2.161 & 1.967 & 2.758 \\
 \hline  
\end{tabular}
\caption{Smooth rising thermal bubble: minimum and maximum horizontal velocity $u$ and vertical velocity $w$ at $t = 600$ s compared
against results reported in \cite{Restelli1}.
}
\label{tab:1}
\end{center}
\end{table}

\subsection{Non-smooth rising thermal bubble} \label{sec:nonsmooth}
The second test case is analogous to the first one, except for a uniform thermal anomaly of $0.5$ K:
\begin{equation}
\theta^0 = 300.5  ~ \textrm{if $r\leq r_c$},\quad\theta^0 = 300
~ \textrm{otherwise},
\label{dcEqn2}
\end{equation}
and $(x_c,z_c) = (500, 260)~\mathrm{m}$.
The time step is set to $\Delta t = 0.1$. Furthermore, we set $\mu = 0.3$ and $Pr = 1$. 

\begin{figure}[htb]
\centering
 \begin{overpic}[width=0.52\textwidth]{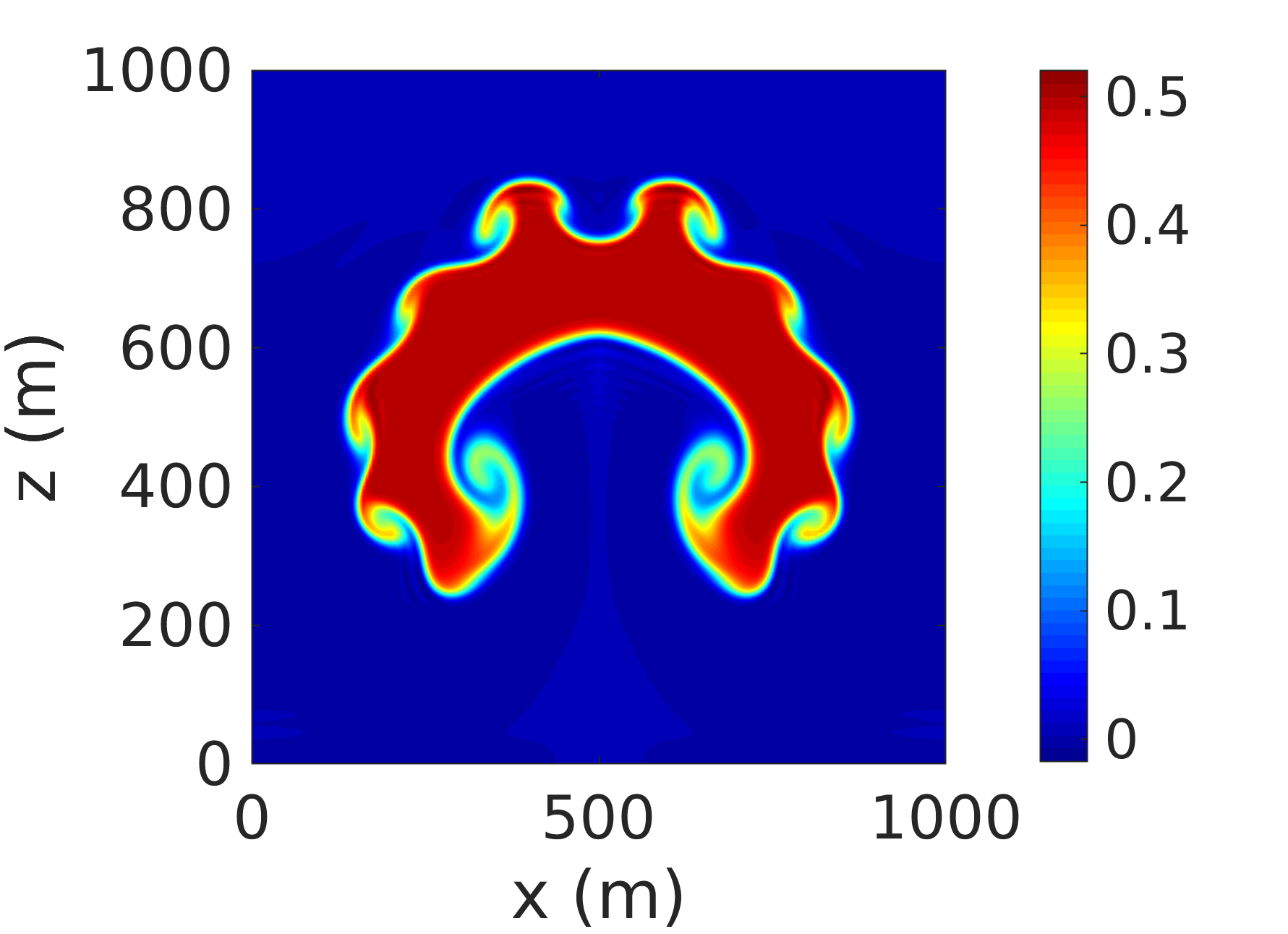}  
      \end{overpic}~ \hspace{0.1cm}
       \begin{overpic}[width=0.52\textwidth]{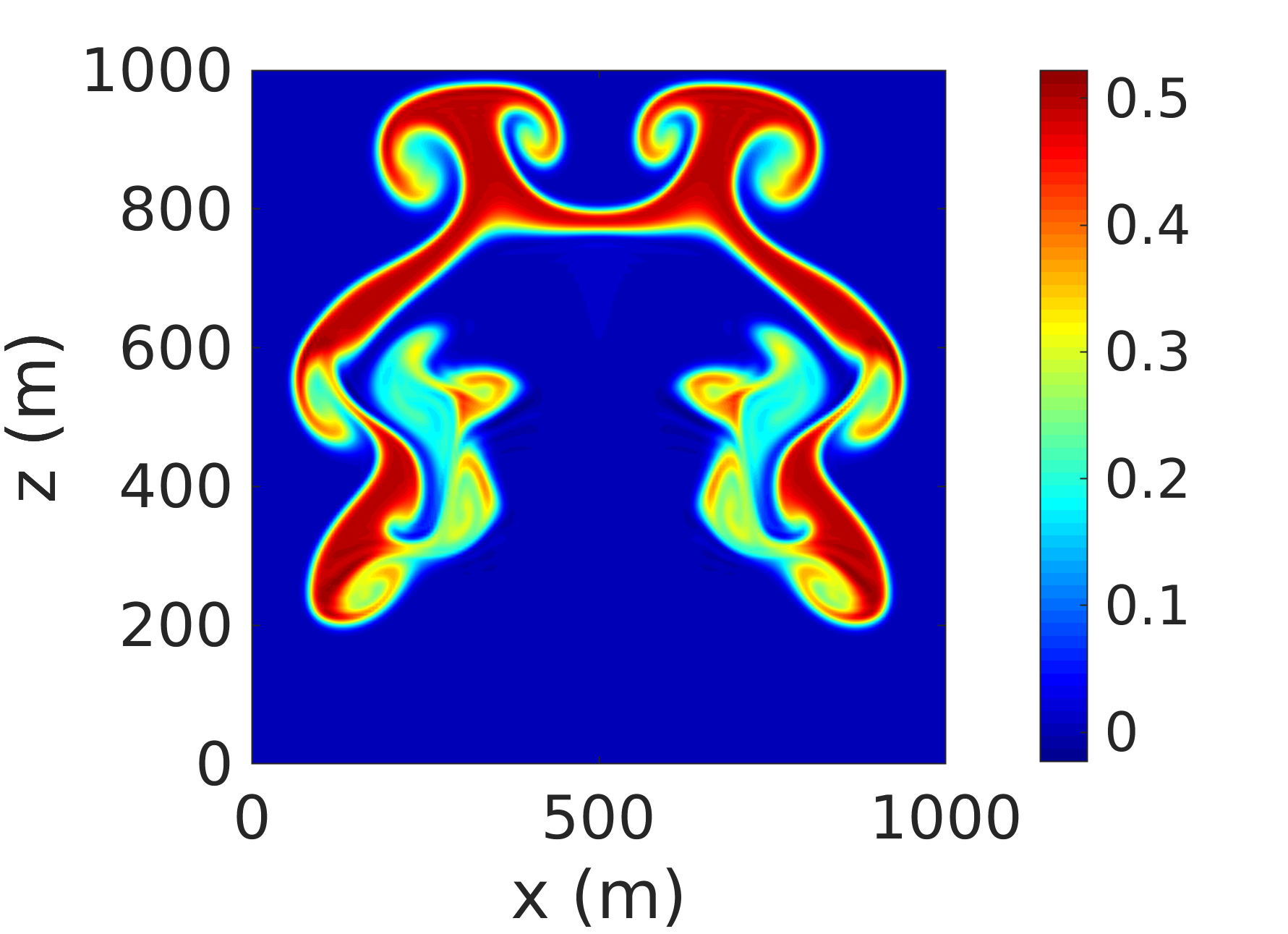}  
      \end{overpic}
\caption{Non-smooth rising thermal bubble: perturbation of potential temperature computed at $t = 420$ s (left) and $t = 600$ s (right).}
\label{fig:RTB4}
\end{figure}

Figure \ref{fig:RTB4} shows the spatial distribution of the potential temperature perturbation
at $t = 420$ s and $t = 600$ s by using a mesh with uniform resolution $h = \Delta x = \Delta z = 5$ m. Qualitatively, these results are in very good agreement
with those reported in \cite{Restelli2}. For a more quantitative analysis, in Table \ref{tab:3} we report the extrema for $\theta'$ at $t = 420, 600$ s together with the values from \cite{Restelli2}. 
We observe that the comparison is satisfactory considering that the method we use is very different from the approach in \cite{Restelli2}.

\begin{table}[htb]
\begin{center}
\begin{tabular}{|c|c|c|c|c|c|} \hline
Model & Res. [m] & $\theta'_{min}$ (K) at $420$ s & $\theta'_{min}$ (K) at $600$ s & $\theta'_{max}$ (K) at $420$ s & $\theta'_{max}$ (K) at $600$ s \\
 \hline
GEA & 5 & -0.018 & -0.023 & 0.521 & 0.522\\
Ref. \cite{Restelli2}  & 5 & -0.020 & -0.016 & 0.522 & 0.527\\
GEA & 10 & -0.047 & -0.062 & 0.577 & 0.563\\
Ref. \cite{Restelli2} & 10 & -0.057 & -0.053 & 0.550 & 0.561\\
 \hline  
\end{tabular}
\caption{Non-smooth rising thermal bubble: minimum and maximum potential temperature perturbation $\theta'$ at $t = 420$ s and $t = 600$ s compared against results reported in \cite{Restelli2}.
}\label{tab:3}
\end{center}
\end{table}

\section{Concluding remarks}
This paper discusses and further assessed GEA (Geophysical and Evinromental Applications), a new C++ language simulation framework
designed for the numerical simulation of atmospheric and ocean flows. We developed a pressure-based solver for the Euler equations written in conservative form using density, momentum, and total
energy as variables. 
We validated the solver against numerical data available in the literature for two well-known
benchmarks: the smooth and non-smooth thermal bubble. For both tests, we obtain good qualitative and quantitative comparisons. 
The code created for this paper is available on GitHub \cite{GEA}.

%
%

\end{document}